\input{aipcheck}

\documentclass[
    ,final            
  ]
  {aipproc}

\layoutstyle{8x11single}

\usepackage{slashed}

\def\be{\begin{equation}}
\def\ee{\end{equation}}
\def\bea{\begin{eqnarray}}
\def\eea{\end{eqnarray}}

\def\gev{\, {\rm GeV}}

\def\kev{\, {\rm keV}}

\newcommand{\gsim}{\lower.7ex\hbox{$\;\stackrel{\textstyle>}{\sim}\;$}}
\newcommand{\lsim}{\lower.7ex\hbox{$\;\stackrel{\textstyle<}{\sim}\;$}}

\newcommand{\ifb}{\rm fb^{-1}}
\newcommand{\pb}{\rm pb}

\newcommand{\cm}{\rm{cm}}

\newcommand{\km}{\rm{km}}

\newcommand{\s}{\rm{s}}

\begin{document}

\title{Novel Dark Matter Models and Detection Strategies}


\author{Jason Kumar}{
  address={Department of Physics and Astronomy, University of Hawaii, Honolulu, HI,
  96822  USA}
}



\begin{abstract}
We consider the impact of relaxing some typical assumptions
about dark matter interactions, including isospin-invariance,
elastic scattering and contact interactions.  We show that
detection strategies with
neutrino detectors, gamma-ray searches, new direct detection
experiments and collider searches can all provide complementary
information.  We argue that data from many such strategies may be necessary
to gain a more complete understanding of dark matter interactions.
\end{abstract}

\maketitle


\section{Introduction}

The idea that experiments could search for dark matter beyond its gravitational
effects was proposed almost 30 years ago~\cite{Goodman:1984dc}.
Since that time, the search for dark matter has become an experimental reality, and dark matter
research is a field which is well beyond its infancy.  Data is arriving from a variety
of detectors using different strategies, and some very interesting hints which may
suggest the presence of dark matter have been seen.

WIMPs of the MSSM are one of the most appealing dark matter candidates, and dark matter search results
are often interpreted with this model in mind.  Aside from the theoretical
appeal of LSP WIMPs, this model is relatively constrained, bringing with it a number
of assumptions which simplify the tasks of interpreting the data and comparing data
from different detectors.  Typical assumptions which one makes include:
\begin{itemize}
\item{{\it Isospin-invariance}: It is assumed that dark matter interactions with protons
and neutrons are the same.}
\item{{\it Elastic interactions}: It is assumed that dark matter scatters elastically off
nuclei.}
\item{{\it Contact interactions}: It is assumed that dark matter-nucleon interactions are
mediated by a very heavy particle.}
\item{{\it Single component}: It is assumed that all dark matter consists of a single
new particle, the MSSM LSP.}
\end{itemize}

But as new data comes in, and potential hints of dark matter are seen by some detectors, it
has becoming increasingly clear that it may be difficult to reconcile the data from all of these
experiments under the assumptions given above.  It is thus necessary to consider how the role
of different detection strategies changes when  the above assumptions are relaxed.\footnote{Dynamical
dark matter~\cite{DDM}, where dark matter is not a single component, is considered elsewhere in these
proceedings.}
We will argue that, in the absence of the above assumptions, it is necessary to
combine the results of several experiments using novel detection strategies in order to study
the nature of dark matter interactions.

\section{Dark Matter Interactions}

It is useful to briefly review the general structure of non-relativistic dark matter-nucleus scattering.
We allow for the possibility that the outgoing dark sector particle may have a mass $m_{X'}$  which is different
from the dark matter mass $m_X$ (we will assume $m_{X'} \geq m_X$).
But since we assume that both incoming and outgoing particles are non-relativistic,
we have $\delta m_X \equiv m_{X'} - m_X \ll m_X$.  We may then write the reduced mass as
$\mu_A = m_X m_A / (m_X +m_A)$ for both the incoming and outgoing system, where $m_A $ is the nucleus
mass.
For cold dark matter, one expects dark matter to coherently scatter off the nucleus as a
whole.  The
dark matter-nucleus differential scattering cross-section is
\bea
{d\sigma \over dE_R } = {\mu_A p_{out} \over 16\pi m_X^2 m_A^2 v }
\left({1\over N} \sum_{spins}|{\cal M}|^2 \right) |F_A (E_R)|^2 {1\over E_+ - E_-}\theta(E_+ - E_R)\theta(E_R - E_-) ,
\eea
where $v$ is the relative velocity of the incoming particles, $p_{out}=\mu_A v \sqrt{1-2\delta m_X /\mu_A v^2}$
is the outgoing momentum in center of mass frame, ${\cal M}$ is the matrix element for dark
matter-nucleus scattering, $N$ is the
number of initial states, and the sum is over all initial and final states.  The nuclear form
factor is given by $F_A (E_R)$, and the maximum (minimum) nucleus recoil energy which is kinematically
possible in two-body scattering is given by $E_+$ ($E_-$):
\bea
E_\pm = {\mu_A^2 v^2 \over m_A } \left(1- {\delta m_X \over \mu_A v^2}
\pm \sqrt{1- {2\delta m_X \over \mu_A v^2}} \right) =
{\mu_A v \over m_A} \left[
{\mu_A v } \left(1- {\delta m_X \over \mu_A v^2} \right)
\pm  {p_{out} } \right] .
\eea
The differential rate of scattering events is then given by~\cite{Gould:1987ir}
\bea
{dR \over dV}= \eta_T \eta_X \int d^3u~f(u) {v^2 \over u}
\int_{E_{min}}^{E_+}dE_R~{d\sigma \over d{E_R}} ,
\label{RateEqn}
\eea
where $\eta_{T,X}$ are the number densities of the target material and of dark matter, respectively.
The dark matter velocity distribution is
given by $f(u)$, where $u$ is the velocity of a dark matter particle relative to the detector when it is
far from the solar system.  Note that $u$ and $v$ can be different, since an infalling dark matter particle will gain speed
before scattering due to the gravitational potential; we then have $(1/2) m_X (v^2 - u^2) = |V_{grav}(r)|$, where
$V_{grav}(r)$ is the gravitational potential energy of the dark matter at $r$.
Usually $u \approx v$, unless scattering occurs within
the sun.  The lower integration limit is given by $E_{min} = \max [E_{th}, E_- ]$,
where $E_{th}$ is the threshold recoil energy.

Thus far we have relied only on kinematics, making no assumptions about dark matter particle physics.
Henceforth, we will focus only on the case where dark matter interactions are spin-independent.  In that
case, the dark matter coupling scales as the number of nucleons and
${\cal M} \propto [f_p Z + f_n(A-Z) ]$, where $f_{p,n}$ parameterize the relative strength of dark
matter interactions with protons and
neutrons, respectively.  If scattering arises through exchange of a single mediating particle with
mass $M_*$, then
we may write the matrix element as
\bea
{\cal M} = {G(m_X , m_A ,s,t,u) \over q^2 - M_*^2} [f_p Z + f_n(A-Z) ],
\eea
where $G$ is a dimension-2 function of the masses and Mandelstam variables.  The momentum transfer in
the hard scattering process, $q$, is give by $q^2 \sim  m_X^2,~-2m_A E_R,~m_X^2$ for $s$-, $t$- and $u$-channel
exchange, respectively.
For $t$-channel exchange, the shape of the recoil spectrum will depend on whether the propagator is dominated
by the messenger mass or the momentum transfer.  Since $|t| < 4 m_A^2 v^2$, scattering interactions
will be short-ranged if $M_* \gsim 1~\gev$.

We will also assume that scattering is velocity-independent, which is the case if
there are no cancelations in the leading order term of the matrix element.
$G$ is determined, up to coupling constants and other numerical factors,  by the normalization
of the incoming and outgoing states and by the derivatives (if any) in the matrix element.
We find that $G \propto m_X m_A$ for fermionic dark matter, or for scalar dark matter which couples
to the mediating particle through a derivative.  For scalar dark matter (examples include~\cite{ScalarExamples})
without a derivative coupling,
we can also have terms of the form $G \propto M_* m_A$.

If dark matter scatters elastically, then we may set $\delta m_X =0$.
If $|t| \ll M_*^2$, then we may write
\bea
{d\sigma^{Z,A} \over dE_R } = {\mu_A^2 \over M_*^4} [f_p Z + f_n (A-Z)]^2 |F_A (E_R)|^2 {\theta(E_+ - E_R) \over E_+}
= {\sigma^p \over E_+ }
{\mu_A^2 \over \mu_p^2}
\left[Z + {f_n \over f_p} (A-Z)\right]^2 |F_A (E_R)|^2 \theta(E_+ - E_R),
\eea
where $\mu_p$ is the dark matter-proton reduced mass and $\sigma^p$ is the dark matter-proton spin-independent
scattering cross-section.  One can
make motivated assumptions about $\eta_X$, $f(u)$ (from astrophysics) and $F_A (E_R)$ (from nuclear physics).
The only parameter left is $f_n / f_p$; given a choice of this parameter, a bound on the event rate can
be directly translated into a bound on $\sigma^p$.  Conversely, an excess event rate can be directly
translated into a preferred region for $\sigma^p$.

\section{Isospin-Violating Dark Matter}

\begin{figure}[tb]
\includegraphics*[width=0.30\columnwidth]{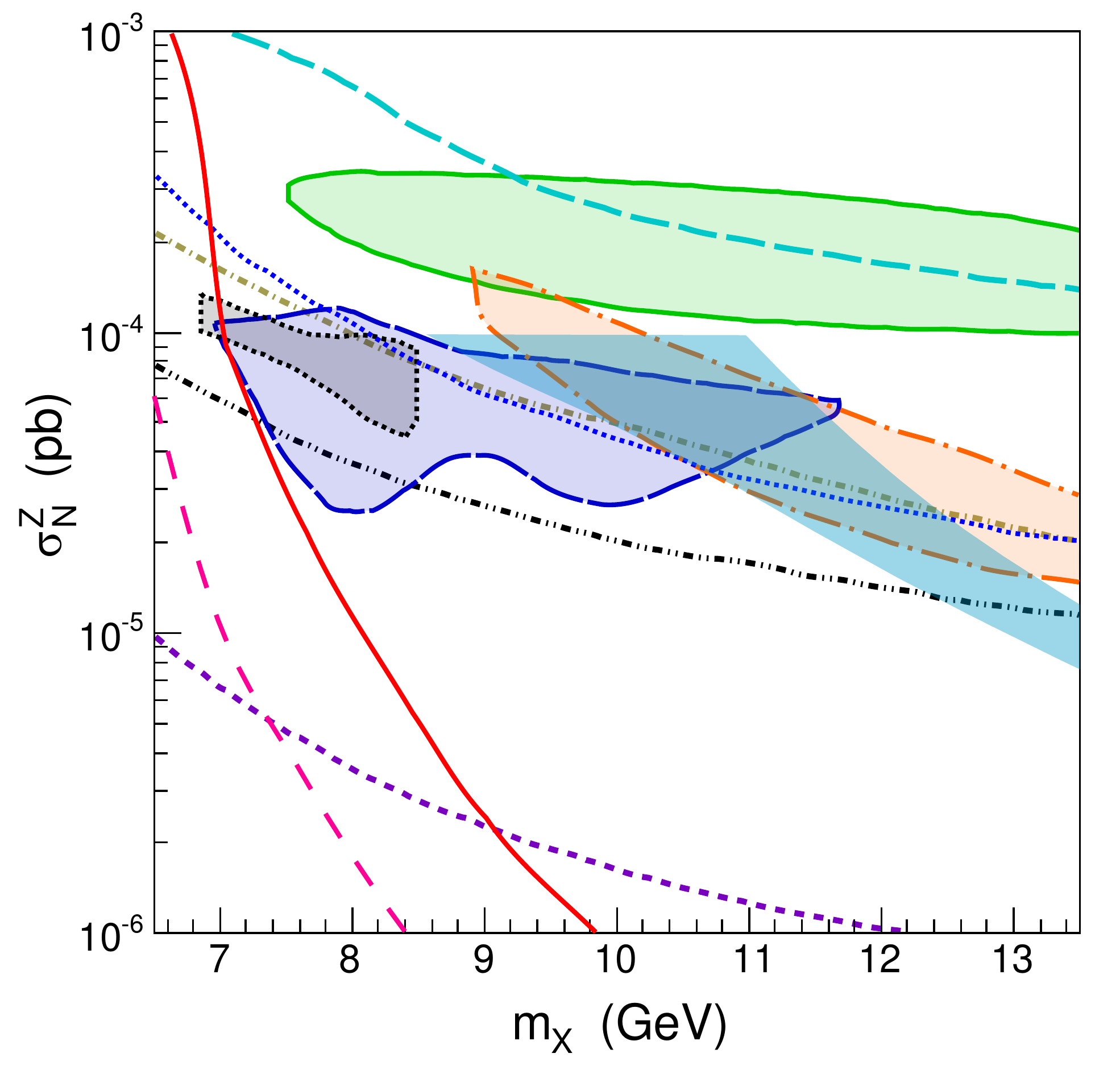}
\includegraphics*[width=0.30\columnwidth]{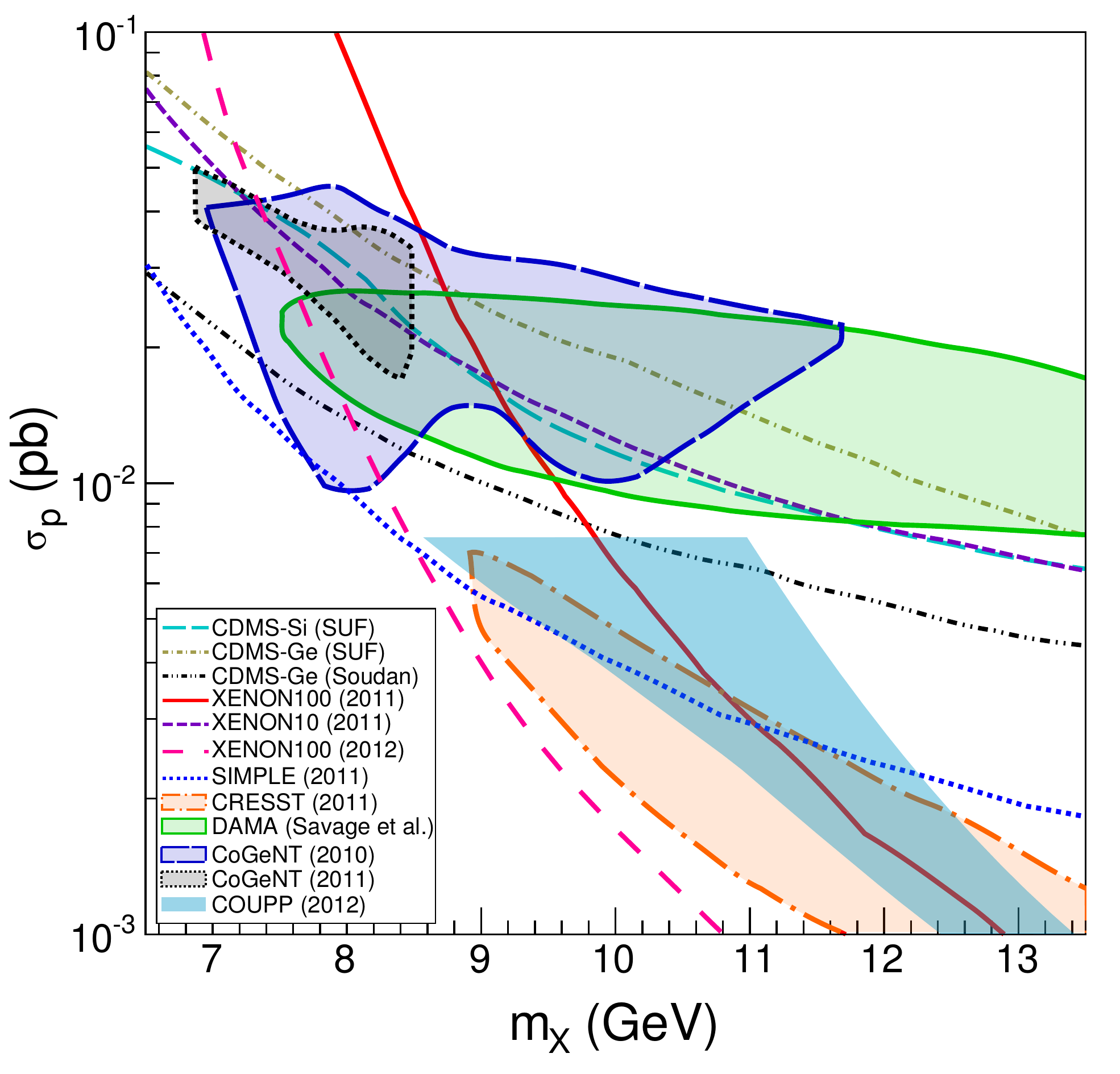}
\vspace*{-.1in}
\caption{\label{fig:DDbounds} Favored regions and exclusion contours
in the $(m_X, \sigma_N^Z)$ plane (left),  and in the
$(m_X, \sigma_p)$ plane for IVDM with $f_n / f_p = - 0.7$ (right).
(Figure courtesy of David Sanford.) }
\end{figure}

Typically, bounds are presented in terms of a ``normalized-to-nucleon" scattering cross-section, $\sigma_N^Z$.
This is the dark matter-nucleon scattering cross-section which would be inferred under the
assumption $f_n = f_p$, and provides an easy way to compare the
results of experiments using different targets.
But the quantity which actually should agree among different experiments is the dark matter-proton scattering
cross-section, $\sigma^p$, which is
related to $\sigma_N^Z$ by the relation
\bea
{\sigma^p \over \sigma_N^Z} \equiv F_Z =
\frac{\sum_i \eta_i \mu_{A_i}^2 A_i^2}
{\sum_i \eta_i \mu_{A_i}^2 [Z + (A_i - Z) f_n/f_p]^2} .
\eea
The sum is over isotopes labeled by $i$, and $\eta_i$ is
the natural abundance of the $i$th isotope (we have assumed that
$F_{A_i} (E_R)$ does not vary much between isotopes).  $F_Z$ thus depends only on
$f_n / f_p$, and as expected, $F_Z = 1$ for $f_n / f_p =1$.  For isospin-violating
dark matter (IVDM)~\cite{olderIVDM,Feng:2011vu}, $f_n / f_p $ is a parameter which is set by the details of the
particle physics model.
Figure~\ref{fig:DDbounds}~\cite{Feng:2011vu} shows the favored regions of DAMA~\cite{Bernabei:2010mq,DAMAregion} ($3\sigma$),
CoGeNT~\cite{Aalseth:2010vx,Aalseth:2011wp} (90\% CL) and CRESST~\cite{Angloher:2011uu} ($2\sigma$),
and $90 \%$ CL exclusion contours from
CDMS~\cite{Akerib:2010pv,Ahmed:2010wy}, XENON10~\cite{Angle:2011th},
XENON100~\cite{Aprile:2010um,Aprile:2011hi,Aprile:2012nq},
SIMPLE~\cite{Felizardo:2011uw} and COUPP~\cite{Behnke:2012ys}.  The left panel plots these regions for $f_n / f_p =1$, while
the right panel assumes $f_n / f_p =-0.7$.

One can see the dramatic effect of isospin-violating interactions
on the relative sensitivity of various direct detection experiments.  For
$f_n / f_p =-0.7$, the CoGeNT and DAMA preferred regions are brought closer
into alignment, while the bounds from XENON10/100 are weakened, and no longer
exclude the entire DAMA and CoGeNT regions.  However, it is clear that this is
not a complete solution.  The marginal tension between bounds from CDMS (Soudan)
and the preferred region of CoGeNT is not altered by isospin-violating interactions,
since both detectors use a germanium target.  Moreover, although the choice
$f_n / f_p =-0.7$ would alleviate the tension with the xenon-based experiments, it
creates tension with bounds from the SIMPLE experiment.  And there is no choice of
$f_n / f_p$ for which the signal regions from DAMA, CoGeNT and CRESST are all
consistent.

There are many experimental uncertainties with the data at low-mass, some of which
will be resolved soon.  CoGeNT is gaining a better understanding of their surface-area
contamination~\cite{Aalseth:2012if}, which will likely move their preferred signal region to higher mass and lower cross-section.
The understanding of the response of sodium and xenon-based detectors to low-energy recoils is improving.
New germanium-based (Majorana, SuperCDMS) and xenon-based (LUX) detectors will soon provide even higher sensitivity
at low-mass.  It seems clear that the possibility of isospin-violating interactions
will have a significant impact on the interpretation of this new data.

If one treats $f_n / f_p$ as a free parameter of the particle physics model, then one finds that
one must use multiple direct detection experiments to get a handle on dark matter interactions.
We may define the ratio
\bea
R[Z_1, Z_2](f_n / f_p) \equiv {F_{Z_2} \over F_{Z_1}} = {\sigma_N^{Z_1} \over \sigma_N^{Z_2}},
\eea
which is the ratio of normalized-to-nucleon cross-sections which would be inferred by two
detectors, using materials with $Z_1$ and $Z_2$ protons, under the assumption of isospin-invariant
interactions.  Signals at two dark matter experiments provide an experimental measurement of
$R$ and can be used to solve for $f_n / f_p$.
If a detector using a material with $Z_1$ protons finds a dark matter signal, $R[Z_1 , Z_2]$ determines the
range of sensitivity a second detector (using a material with $Z_2$ protons) would need to either potentially
confirm or definitively exclude this signal for any choice of $f_n / f_p$.

If an element has multiple isotopes, no choice of $f_n / f_p$ can
result in total destructive interference between proton and neutron couplings for all isotopes.
For example, marginalizing over $f_n / f_p$, we find
$\max \{R[Z_1= Ge , Z_2 = Xe] (f_n / f_p)\} \sim 22$; a xenon-based detector would need at most a factor of 22
greater sensitivity to confirm a signal at a germanium-based detector, for any choice of $f_n / f_p$.  Data
from LUX could thus provide a much more definitive test of the signals from DAMA, CoGeNT and CRESST, assuming
the low-energy response can be well understood.

\begin{figure}[tb]
\includegraphics[scale=0.35]{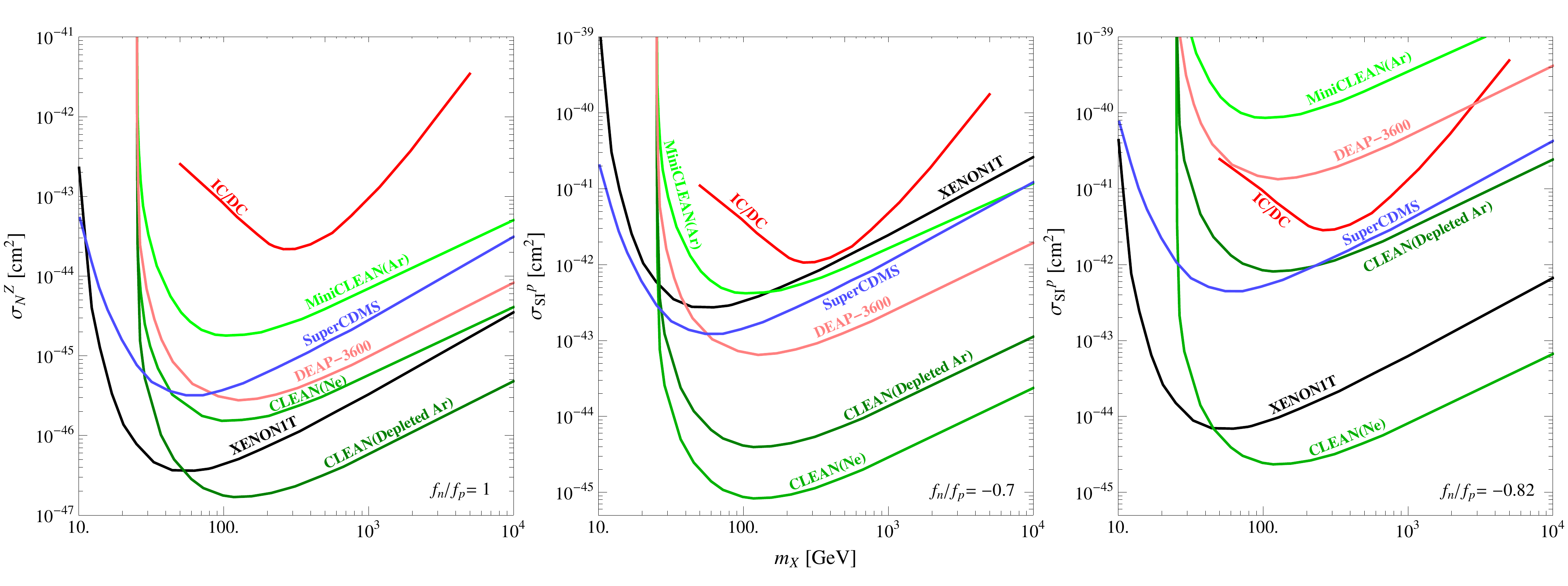}
\vspace*{-.1in}
\caption{\label{fig:sigmaSIoptimalfuture}
Sensitivity to $\sigma^p$ for  $f_n/f_p=1$ (left panel),
$f_n/f_p=-0.7$ (center panel) and $f_n/f_p=-0.82$ (right panel) for
IC/DC with 180 days of data, and for other labeled experiments (see text).
}
\end{figure}

Although IVDM is often discussed in relation to low-mass dark matter, it is a possibility which one must
consider for any dark matter signal.  Figure~\ref{fig:sigmaSIoptimalfuture}~\cite{Gao:2011bq} shows the sensitivities of
XENON1T~\cite{bib:xenon100}, Super-CDMS~\cite{SuperCDMSlimit}, MiniCLEAN, DEAP-3600 and CLEAN~\cite{DEAPCLEAN},
for $f_n / f_p = 1, -0.7,~{\rm or}~-0.82$.
Also plotted in figure~\ref{fig:sigmaSIoptimalfuture} is the estimated sensitivity of IceCube/DeepCore~\cite{bib:ic80dc} with
180 days of data (assuming annihilation to the hard channel).

\subsection{Dark Matter Searches with Neutrino Detectors}

IceCube/DeepCore and other neutrino detectors search for the neutrino flux arising from dark matter
annihilation in the core of the sun.  If the sun is in equilibrium, bounds on the dark matter annihilation
rate can be directly translated into bounds on the rate at which dark matter is captured by the sun through
scattering off solar nuclei.\footnote{This bound also depends on the neutrino spectrum produced by dark matter
annihilation; we assume annihilation only to $\bar \tau \tau$ for $m_X \leq 80~\gev$, and annihilation
only to $W^+ W^-$ for $m_X > 80~\gev $.}  In turn, this yields a bound on the dark matter-proton scattering cross-section.
The dark matter capture rate can also be derived from eqn.~\ref{RateEqn}, where the threshold energy $E_{th}$
is the minimum recoil energy needed for dark matter to be captured.

Dark matter capture occurs when a dark matter particle scatters and loses enough energy to become confined to
an orbit around the sun.  After many subsequent scatterings, it will eventually settle to the core of the sun, where it
annihilates.  But
since many-body effects can be important for dark matter in large-radius orbits, one often requires captured dark matter
to be confined to an orbit which will not exceed a maximum distance $r_0$ from the sun (typically taken to be the
Jupiter-sun distance).  The minimum recoil energy necessary for capture is thus given by
\bea
E_{th} = \max \left[ {1\over 2}m_X (u^2 + v_{esc.} (r_0)^2) -\delta m_X , 0 \right] ,
\eea
where $v_{esc.} (r_0)$ is the escape velocity for a particle at distance $r_0$ from the sun.

One can then determine the capture rate from scattering against each element by integrating the differential
capture rate (eq.~\ref{RateEqn}) throughout the volume of the sun, accounting for the densities of different elements, as well
as the sun's gravitational potential.
If one computes the capture rate for each isotope of each element under the assumption of isospin-invariant
interactions, then one obtains the capture rate for any choice of $f_n / f_p$ by simply rescaling by the
factor $[Z +(f_n / f_p)(A-Z)]^2 / A^2$.

As shown in figure~\ref{fig:sigmaSIoptimalfuture},
when there is destructive interference between proton and neutron scattering, neutrino detectors
can provide a nice complementary probe with a sensitivity which is comparable to that of direct detection
experiments.  This is because many direct detection experiments use heavy nuclei which have many more neutrons
than protons; scattering in the sun is largely off lighter elements (including hydrogen), with fewer neutrons.

Neutrino detectors can also provide sensitivity to low-mass dark matter which is comparable to that of
direct detection experiments~\cite{LowMassNeutrinoDetector,Kumar:2011hi,Kumar:2012uh}.  This can easily be understood from
eq.~\ref{RateEqn}.  For $m_X \sim 5-10~\gev$ many direct detection experiments satisfy $m_X \ll m_A$,
$E_+ \sim 2 m_X^2 v^2 /m_A$, $E_{th}={\rm fixed}$, implying that the event rate vanishes for
small $m_X$.  For neutrino detectors, however, there are
many relatively light elements with mass comparable to the dark matter, implying $E_+ \sim 2m_X v^2$, while
$E_{th} = (1/2) m_X u^2 < E_+$.  In fact, neutrino detectors can be sensitive to dark matter as light as
$4~\gev$; lighter dark matter will evaporate out of the sun~\cite{Gould:1987ir}.

We will consider a search for dark matter annihilation utilizing a 1 kT liquid scintillation (LS) detector
with 2135 live-days of data (roughly the specifications of KamLAND).
Neutrino detectors search for the charged lepton produced from $\bar \nu, \nu$ by a charged-current interaction.
It has recently been realized that
LS detectors can also be used for dark matter searches, because the direction of the charged lepton track
in the scintillator can be determined from the timing of the first photons reaching the PMTs~\cite{LSDetector}.  Moreover, the charged
lepton flavor can be determined with very high efficiency.  For low-energy $\nu_\mu$ (arising from low-mass
dark matter annihilation), the muons produced  by a charged-current interaction will be short-ranged; there is thus
no great advantage in searching for muons which may have been produced outside the detector volume.  Instead, we focus on
a search for $e^\pm$ produced from electron (anti-)neutrinos, which have the advantage of a much smaller
atmospheric neutrino background.

\begin{figure}[tb]
\includegraphics*[width=0.4\columnwidth]{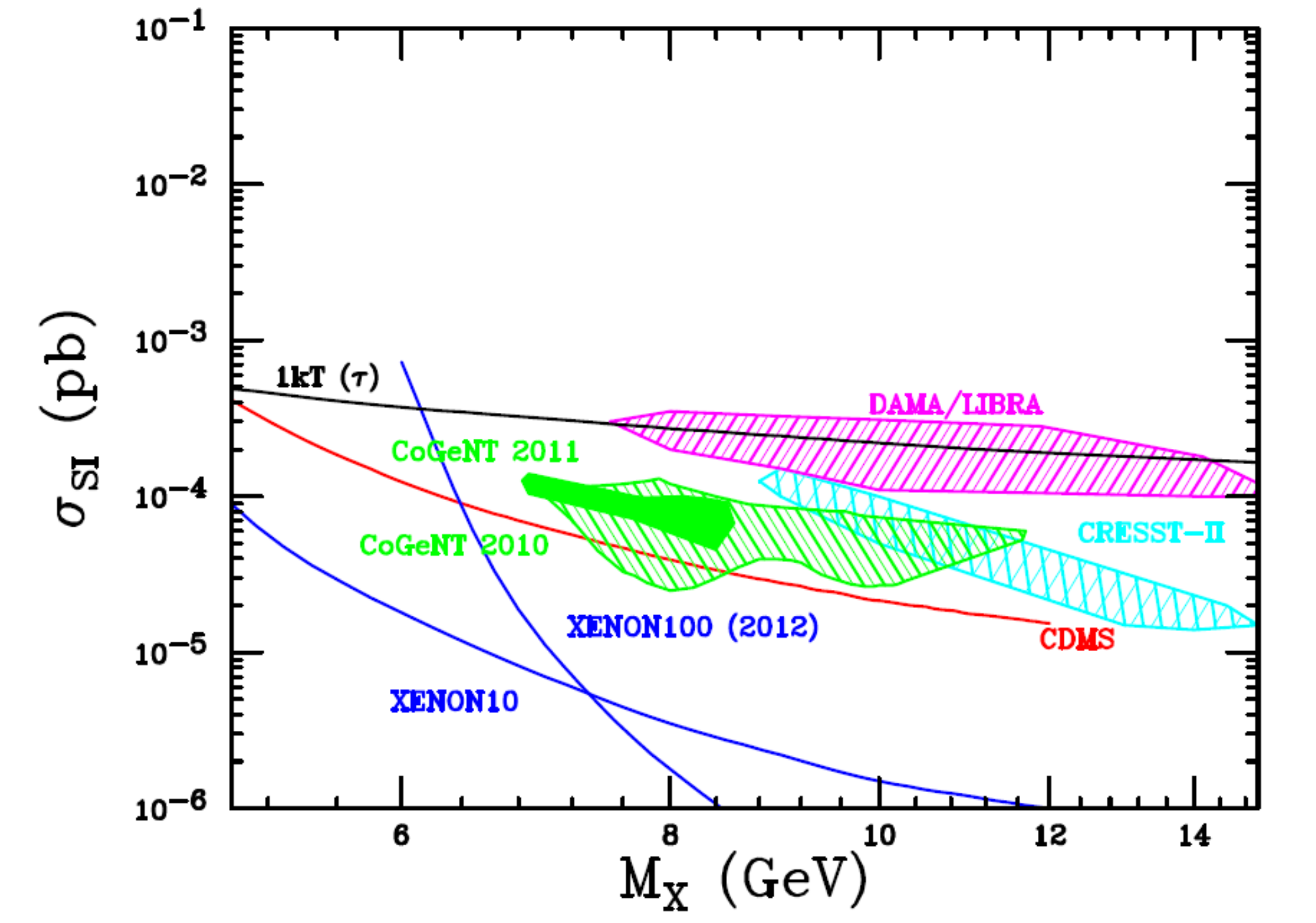}
\includegraphics*[width=0.4\columnwidth]{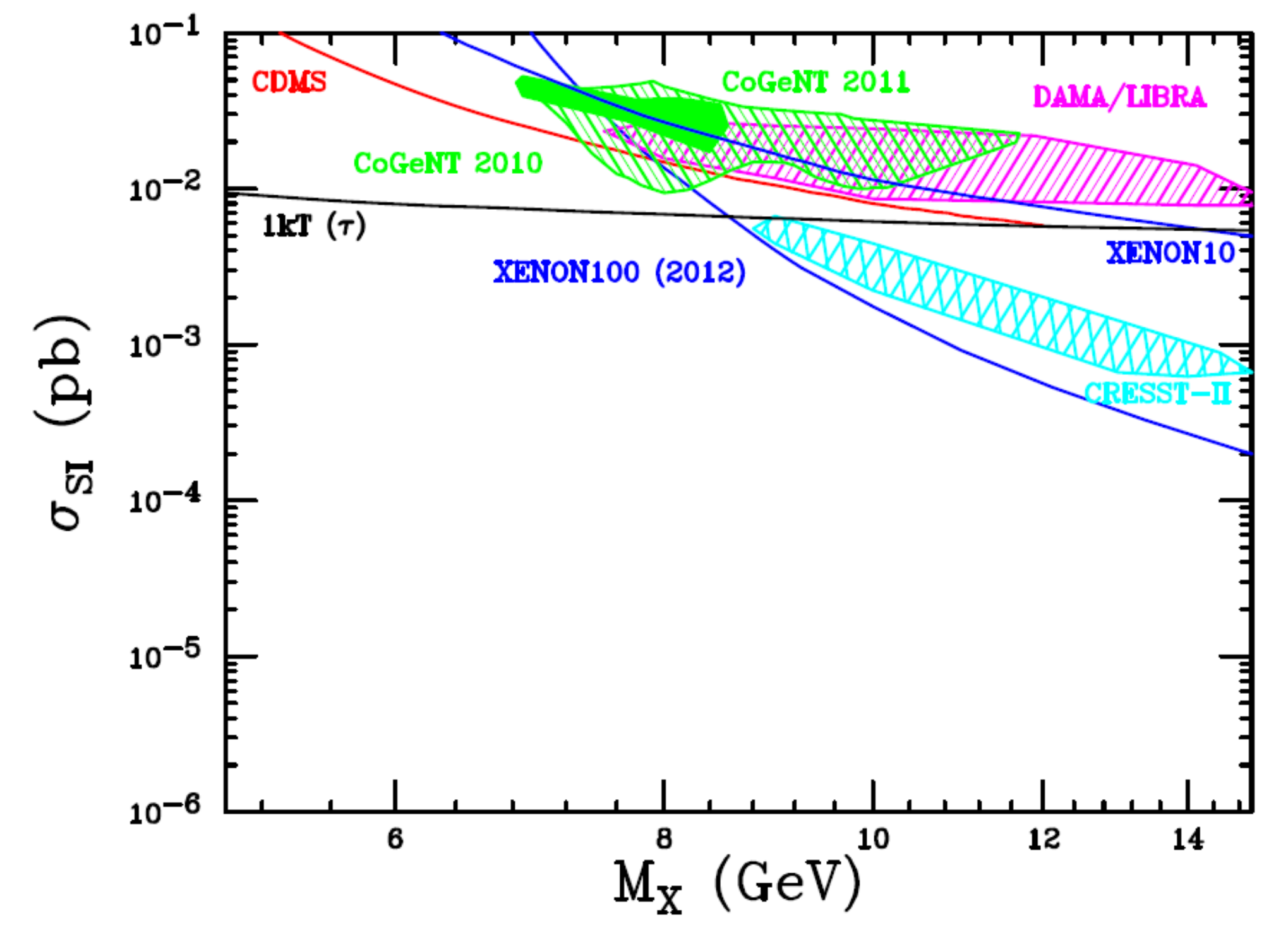}
\vspace*{-.1in}
\caption{\label{fig:IVDMKamLAND} Favored regions and exclusion contours
in the $(m_X, \sigma_p)$ plane for $f_n / f_p =1$ (left)  and
for IVDM with $f_n / f_p = - 0.7$ (right).
(Figure courtesy of Stefanie Smith.) }
\end{figure}

In figure~\ref{fig:IVDMKamLAND}~\cite{Kumar:2011hi} we plot the sensitivity
which can be achieved with KamLAND's current data set (assuming
annihilation to $\bar \tau \tau$), as
well as the DAMA, CoGeNT and CRESST signal regions and the CDMS, XENON10 and XENON100 exclusion curves, assuming
$f_n / f_p =1,-0.7$.  We can see that KamLAND, with its current data set, can potentially probe the signals of
DAMA and CoGeNT if $f_n / f_p \sim -0.7$.

\subsection{Complementary Bounds from Indirect Detection and the LHC}

Indirect detection and collider bounds provide interesting complementary
tests of IVDM models.
The $Xq \rightarrow Xq$ scattering matrix element is related to the $XX \rightarrow \bar q q$
annihilation matrix element and the $\bar q q \rightarrow XX$ production matrix element by
crossing symmetry.  As a result, if an assumption is made about the form and flavor dependence
of the scattering matrix
element, then bounds on the total annihilation cross-section $\langle \sigma_A v \rangle$  or
on the production cross-section can be directly translated
into a bound on $\sigma^p$.
The sensitivity of direct detection experiments using heavy targets, such as xenon,
is minimized for the case of partial destructive interference between $f_n$ and $f_p$.
In that case, a signal
seen at CoGeNT corresponds to a large value of $\sigma^p$ (${\cal O}(10^{-2})~\pb$).
This implies a large coupling to up- and down-quarks, which destructively interfere.
The dark matter annihilation cross section to up- and down-quarks, or production cross-section
from $\bar q q$ initial states, can then be greatly enhanced.
A key point to note, though, is that the kinematics of these complementary
processes can be different.  For the annihilation process, the momentum transfer
is roughly $2m_X$, while for the production process, the momentum transfer is typically $> 2m_X$.

We will consider the case where dark matter-nucleus scattering is spin- and velocity-independent,
and arises from tree-level $t$-channel exchange with quarks.  Furthermore, we will only consider
the case where the resulting annihilation matrix element is not $p$-wave suppressed.  For scalar
dark matter (real or complex), these conditions are only satisfied by exchange of a scalar, while
for Dirac fermion dark matter one must exchange a vector (these conditions can thus not be satisfied
for Majorana fermion dark matter).
In Table~\ref{table:operators},  we list the energy dependence of the scattering, annihilation and
creation matrix elements, as well as the effective scattering operator in the limit where
scattering is a contact interaction ($|t| \ll M_*^2$).
If we assume that dark matter couples only to up- and down-quarks in the limit of elastic contact
scattering, then for any given choice of the contact operator and $f_n / f_p$, we can directly relate the
annihilation  and production cross-sections to the scattering cross-section.

\subsection{Fermi-LAT Searches of Dwarf Spheroidals}

Tight bounds can be placed on dark matter annihilation in dwarf spheroidal galaxies with
data from the Fermi-LAT~\cite{Ackermann:2011wa,GeringerSameth:2011iw}.  These bounds can be phrased in terms of the quantity
\bea
\Phi_{PP} &\equiv& {\langle \sigma_A v \rangle \over 8\pi m_X^2}
\int_{E_{thr}}^{m_X}\sum_f  B_f {dN_f \over dE} dE ,
\label{BoundEqn}
\eea
where $\sigma_A$ is the total annihilation cross-section and $B_f$ and $dN_f / dE$ are
the branching fraction and photon spectrum, respectively, for annihilation to final state $f$.
For a threshold energy $E_{thr} = 1~\gev$, the 95\% CL bound is
$\Phi_{PP} \leq 5.0^{+4.3}_{-4.5} \times 10^{-30} \cm^3 \s^{-1} \gev^{-2}$~\cite{GeringerSameth:2011iw}, where the
uncertainties in the bound arise from systematic uncertainties in the density profile of
the dwarf spheroidal galaxies.

\begin{figure}[tb]
\includegraphics*[width=0.55\columnwidth]{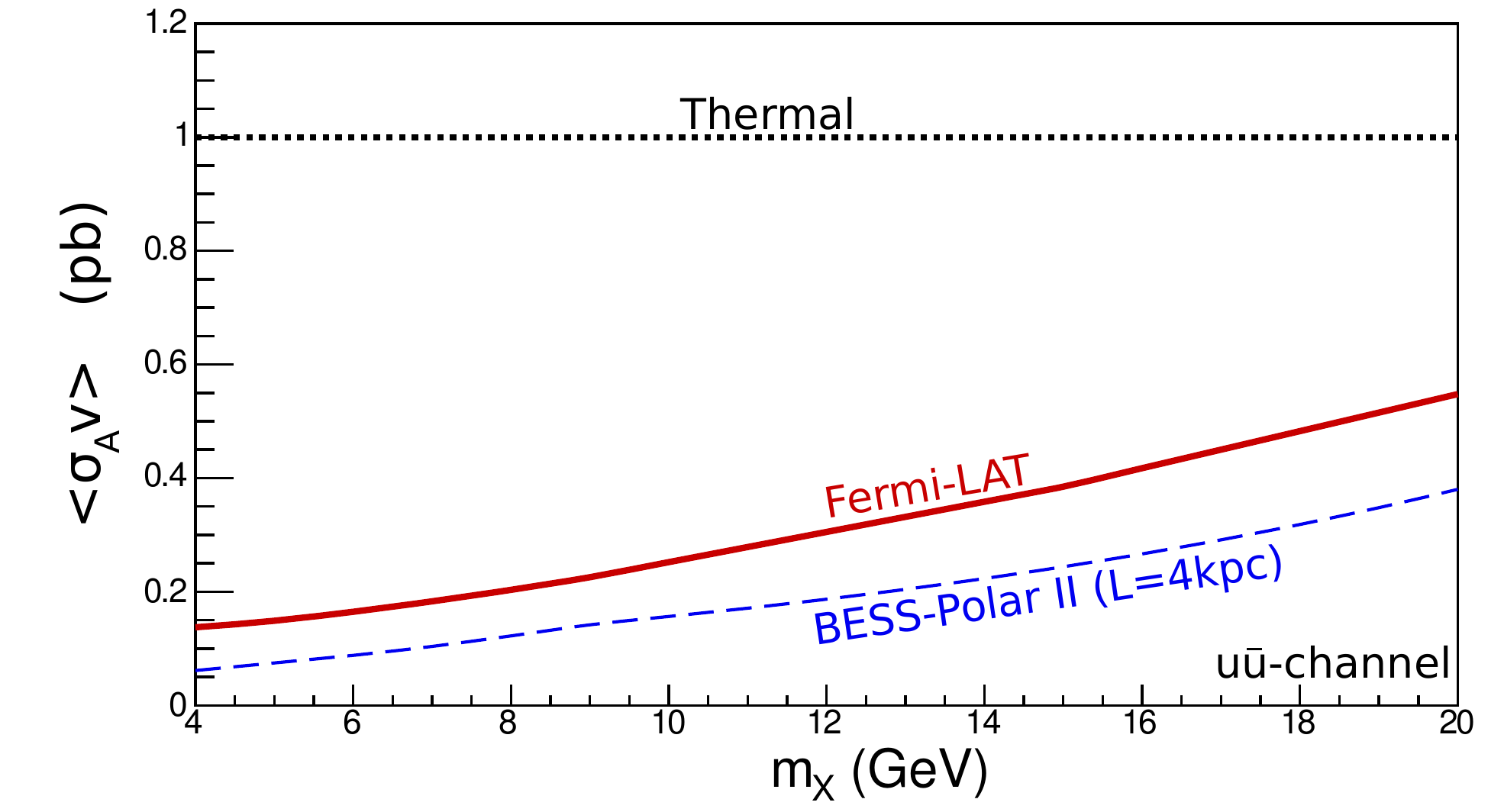}
\includegraphics*[width=0.35\columnwidth]{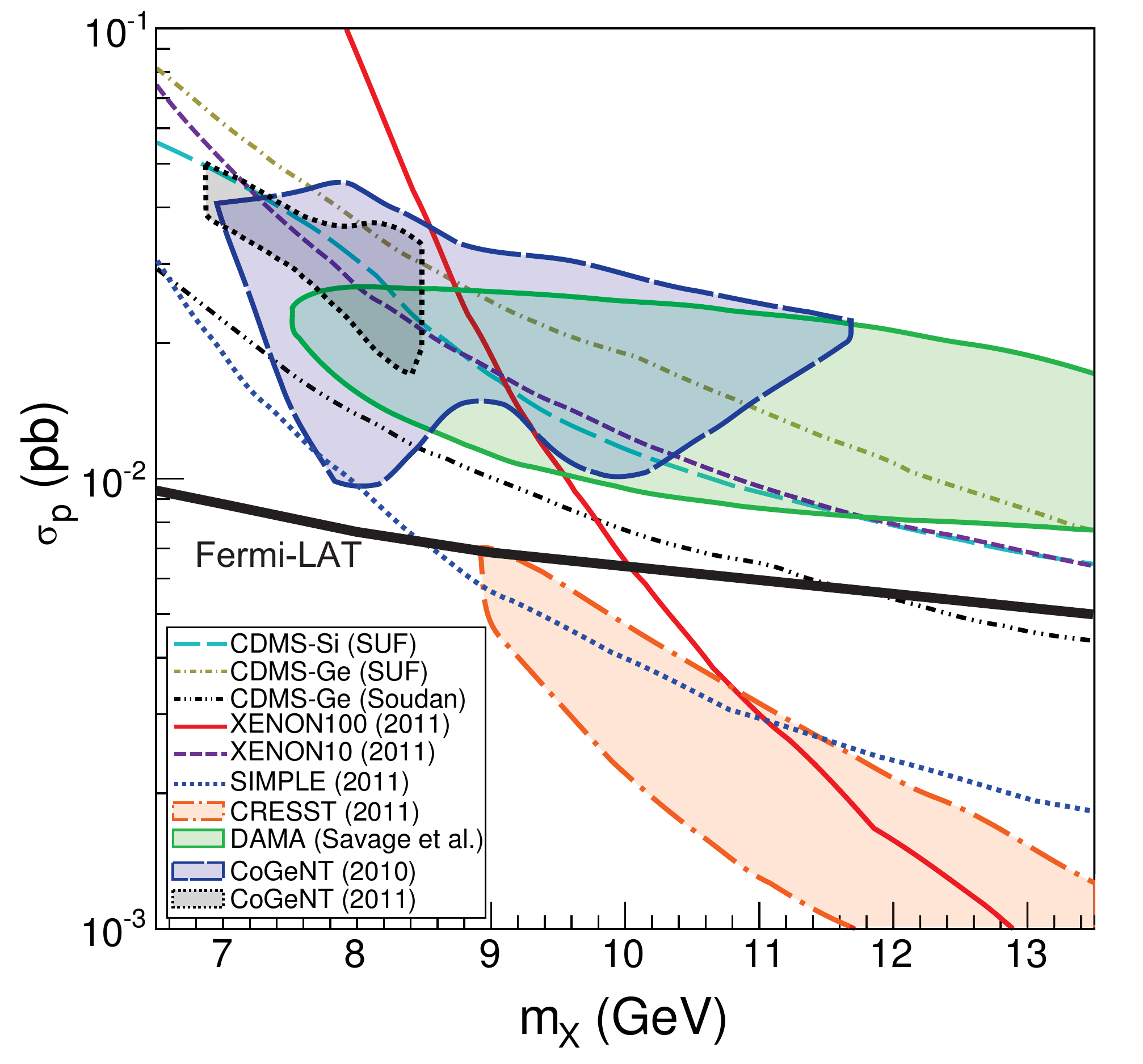}
\vspace*{-.1in}
\caption{\label{fig:IVDMFermi} The left panel shows bounds on the annihilation cross-section
$\langle \sigma_A v \rangle $ from Fermi data and from BESS-Polar II.  The dashed line labeled
``Thermal"  is at $\langle \sigma_A v \rangle =1~\pb$.  The right panel shows
favored regions and exclusion contours for the labeled experiments
in the $(m_X, \sigma_p)$ plane for IVDM with $f_n / f_p = - 0.7$.
The solid black line is the 95\% CL bound from Fermi-LAT data.}
\end{figure}

We computed $dN_f / dE$ for low-mass dark matter annihilation to the up- and down-quark channels
(the two channels are essentially identical) using the Hawaii Open Supercomputing Center (HOSC).
The corresponding bounds on $\langle \sigma_A v \rangle$ are plotted in the left panel of
figure~\ref{fig:IVDMFermi}~\cite{Kumar:2011dr}, along with bounds from BESS-Polar II measurements of the anti-proton
flux~\cite{arXiv:1110.4376}.  Systematic uncertainties can weaken these Fermi bounds by up to a factor of 2, or strengthen
them by up to a factor of 10.  In contrast, BESS-Polar II bounds may be weakened by up to a factor
of 50~\cite{Evoli:2011id} due to systematic uncertainties in cosmic ray propagation, solar modulation, etc.

These bounds can be directly translated into bounds on $\sigma^p$, under the assumption
$|t|, m_X^2 \ll M_*^2$.
For the case when dark matter is a complex
scalar ($f_n / f_p \sim -0.7$), these bounds are plotted in the right panel of figure~\ref{fig:IVDMFermi}.
If dark matter is a real scalar or Dirac fermion, then these bounds are tightened by a factor
of $2$ or $\sim 4$, respectively.

\begin{table}[h]
\caption{ Contact operators (arising from $t$-channel exchange) which permit spin-independent, velocity-independent scattering and
$s$-wave annihilation.  Also listed are the energy dependence of the scattering, annihilation and production matrix elements.}
\begin{tabular}{|c|c|c|c|c|c|}
  \hline
  particle & exchange & contact operator & ${\cal M}_{Xq\rightarrow Xq} $ & ${\cal M}_{XX \rightarrow \bar q q } $ &
  ${\cal M}_{\bar q q \rightarrow XX} $ \\
  \hline
  complex scalar & scalar & ${\cal O}_C=(1/M_*) X^* X \bar q q$ & $\propto {M_* m_A \over t-M_*^2}$ & $\propto {M_* m_X \over 4m_X^2 -M_*^2}$ &
  $\propto {M_* \sqrt{s_{prod.}} \over s_{prod.} -M_*^2}$ \\
  real scalar & scalar & ${\cal O}_R=(1/M_*) X^2 \bar q q$ & $\propto {M_* m_A \over t-M_*^2}$ & $\propto {M_* m_X \over 4m_X^2 -M_*^2}$ &
  $\propto {M_* \sqrt{s_{prod.}} \over s_{prod.} -M_*^2}$ \\
  Dirac fermion & vector & ${\cal O}_D=(1/M_*^2) \bar X \gamma^\mu X \bar q \gamma_\mu q$ & $\propto {m_X m_A \over t-M_*^2}$ &
  $\propto {m_X^2 \over 4m_X^2 -M_*^2}$ & $\propto {s_{prod.} \over s_{prod.} -M_*^2}$ \\
  \hline
\end{tabular}
\label{table:operators}
\end{table}

We can see that these constraints would rule out an IVDM model which could match the DAMA and CoGeNT data.
But this is subject to the specific assumptions which we have made.  For example, if
dark matter couples to up- and down-quarks through an elastic contact operator which permits
spin-independent, velocity-independent scattering, but only $p$-wave suppressed annihilation, then
models which could potentially match the DAMA and CoGeNT data would be unconstrained by Fermi bounds.
Another possibility is that dark matter may couple to Standard Model matter through a relatively
light mediator with mass $M_* \sim 1~\gev$.  In this case, scattering interactions would
still be short-ranged, but dark matter annihilation would have an
additional $M_*^4 / (2m_X)^4 \ll 1$ suppression.

\subsection{Monojet searches at the LHC}

An interesting way to bound direct dark matter production is with monojet or monophoton searches
at colliders~\cite{Aad:2011xw,MonojetCollider}.
Much work has been done on this search strategy~\cite{monojet}.
The key uncertainties in setting this bound arise from the flavor structure of dark matter-quark couplings,
and the energy dependence of the $\bar q q \rightarrow XX$ matrix element.  For example, if one assumes that the dark matter coupling
to quarks is proportional to the quark mass, then a large contribution to the LHC event rate
will typically come from couplings to $\bar c c$, $\bar b b$, even though these couplings yield 
a relatively small contribution
to the scattering cross-section.

Similarly, if the matrix element scales with negative powers of the energy,  the
dark matter production cross-section at the LHC will also be suppressed, since the energy scale of the dark matter production process
is $> 2m_X$ due to the phase space suppression at the production threshold.  Fermionic dark matter is tightly
constrained by LHC monojet/monophoton bounds, since the relativistic normalization of the dark matter state
will introduce additional positive powers of the energy of the process, as seen in Table~\ref{table:operators}.
On the other hand, models with $M_* \sim 1~\gev$ will have negative powers of the process energy from the propagator, and are very weakly
constrained.

\begin{table}
\begin{tabular}{ccc}
\begin{tabular}{|r|r|r|r|}
    \hline
    $m_X$ & ${\cal O}_{D}$ & ${\cal O}_{C}$ & ${\cal O}_{R}$    \\
    \hline
    $4~\gev$ & $0.00285~\pb $ & $90.6~\pb $ & $181~\pb $ \\
    $7~\gev$ & $0.00320~\pb $ & $35.4~\pb $ & $70.3~\pb $ \\
    $10~\gev$ & $0.00357~\pb $ & $18.9~\pb $ & $37.6~\pb $ \\
    $15~\gev$ & $0.00370~\pb $ & $9.1~\pb $ & $18.2~\pb $ \\
    $20~\gev$ & $0.00380~\pb $ & $5.4~\pb $ & $10.9~\pb $ \\
    \hline
    \end{tabular}
    & \qquad\qquad &
    \begin{tabular}{|r|r|r|r|}
    \hline
    $m_X$ & ${\cal O}_{D}$ & ${\cal O}_{C}$ & ${\cal O}_{R}$    \\
    \hline
    $4~\gev$  &  0.00079  &  $10.8~\pb $  &  $21.6~\pb $ \\
    $7~\gev$  &  0.00092  &   $4.2~\pb $  &  $8.50~\pb $ \\
    $10~\gev$ &  0.00097  &   $2.3~\pb $  &  $4.51~\pb $ \\
    $15~\gev$  &  0.00106  &  $1.1~\pb $  &  $2.13~\pb $ \\
    $20~\gev$  &  0.00107  &  $0.62~\pb $  &  $1.24~\pb $ \\

    \hline
    \end{tabular} \\
\small{(a) $p_T > 120~\gev$} & & \small{(b) $p_T > 350~\gev$} \\
\small{\ \ \ \ $\slashed{E}_T > 120~\gev$} & & \small{\ \ \ \ $\slashed{E}_T > 300~\gev$}
\end{tabular}
\caption{Upper bounds on $\sigma^p$ from ATLAS monojet
  searches, assuming the listed cuts on the leading jet $p_T$ and on
  the missing transverse energy.  The
  columns correspond to Dirac fermions, complex scalars, and real
  scalars respectively.}
\label{table:collider}
\end{table}

In Table~\ref{table:collider}~\cite{Kumar:2011dr}, we indicate bounds on $\sigma^p$ derived from
ATLAS monojet searches ($pp \rightarrow XXj$) with $1~\ifb$ of data~\cite{Aad:2011xw}, if we assume
dark matter couples only to first generation
quarks through operators which permit elastic contact spin-independent velocity-independent
scattering and $s$-wave annihilation.  We assume $f_n / f_p =-0.7$.
These bounds arise from analyses using two possible cuts on $\slashed{E}_T$ and on
the $p_T$ of the leading jet, as listed.
The bounds
on scalar dark matter are not competitive, but bounds
on Dirac fermion dark matter can be competitive with (and in some cases exceed) bounds
from direct detection experiments and Fermi.  However, if dark matter interacts though a low-mass 
mediator ($M_* \lsim 1~\gev$), then bounds from the LHC will be weakened by several orders
of magnitude.

\section{Long-Range Interactions}

If dark matter scatters through a very low-mass mediator ($M_*^2 \ll 2m_A E_{th}$), then the
differential scattering cross-section scales
like $(2m_A E_R)^{-2}$ instead of $M_*^{-4}$.  As a result, light elements will tend to have enhanced
scattering cross-sections.  For example, the event rate at neutrino detectors is expected to
receive a significant enhancement relative to leading direct detection experiments, due to the presence
of many light elements in the sun.

In the case of elastic long-range interactions, unlike elastic contact scattering, we cannot
parameterize bounds on dark matter-nucleon scattering in terms of the total cross-section, which is
infinite (an example is Rutherford scattering).
Instead, we can parameterize the dark matter-nucleus differential scattering cross-section as:
\bea
{d\sigma^{Z,A} \over dE_R} &=& C {4\pi \alpha^2 \mu_A^2 \over m_A^2 E_R^2 E_+}
\left[Z + {f_n \over f_p}(A-Z)\right]^2 |F_A(E_R)|^2 ,
\label{LongRangeScatteringEq}
\eea
where $C$ is a constant which normalizes the strength of the mediator coupling to dark matter ($g_X$) and
to the proton ($g_p$) to the proton charge; $C = g_X^2 g_p^2 / e^4$.

Figure~\ref{fig:LongRange}~\cite{Kumar:2012uh} shows the
sensitivity of a 1 kT LS detector with 2135 days of data, assuming annihilation exclusively to either the
$\tau$, $b$, $c$, $\nu$ (flavor-independent) or $g$ channels.  The neutrino spectra arising from dark
matter annihilation to these channels was simulated using the HOSC cluster.
We also plot the estimated
sensitivity of CDMS, if their
current bounds are reinterpreted as bounds on long-range dark matter interactions.  Note that the CDMS
sensitivity is derived under the assumption that the detector's efficiency is constant above the threshold;
as a result of these and other assumptions~\cite{Kumar:2012uh}, the CDMS sensitivity curve should only be regarded as an estimate.
Nevertheless, it is clear that neutrino detectors have a sensitivity to dark matter models with long-range
interactions which can rival leading direct detection experiments.  A detector with a 51 kT liquid argon
target~\cite{Akiri:2011dv}
could achieve the same sensitivity as that given in figure~\ref{fig:LongRange} with $\sim 17$ days of data.

\begin{figure}[tb]
\includegraphics*[width=0.55\columnwidth]{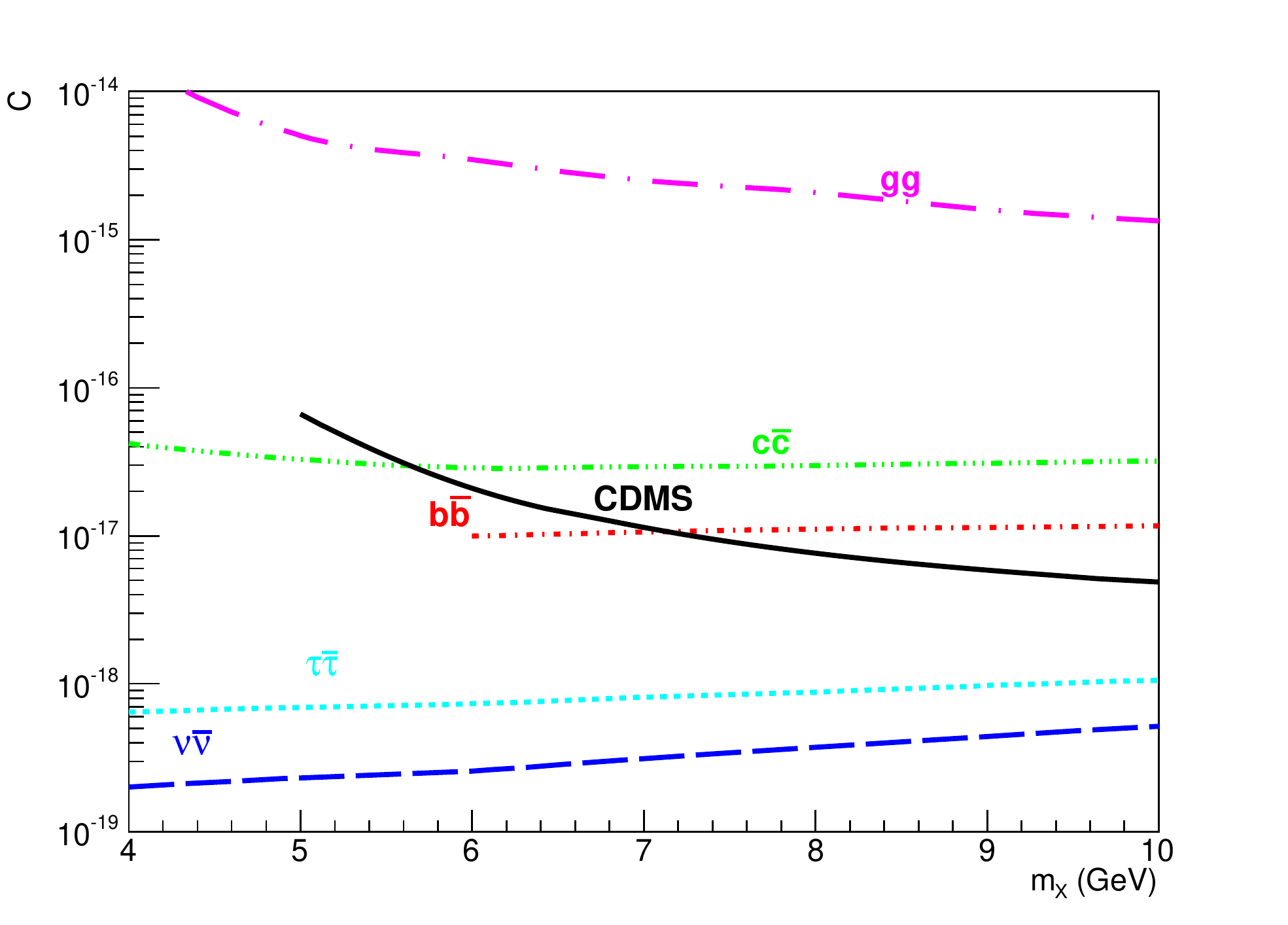}
\vspace*{-.1in}
\caption{\label{fig:LongRange} Sensitivity to $C$  of CDMS and a 1 kT LS detector (2135 live days of data) for
low-mass dark matter with isospin-invariant elastic long-range interactions.  LS detector
sensitivity is shown assuming annihilation to either the $\tau$, $b$, $c$, $g$ and $\nu$
(flavor-independent) channels.  }
\end{figure}

\section{Inelastic Dark Matter (iDM)}

If dark matter scattering is inelastic~\cite{iDM}, then neutrino detectors can provide interesting detection
possibilities which are complementary to direct detection experiments.  For iDM, we
again cannot characterize the bounds from experiments in terms of the dark matter-nucleon scattering
cross-section, because there are kinematic regimes where dark matter-nucleon scattering
is forbidden, while dark matter can still scatter off heavier elements.  The reason is easy to see; for inelastic
scattering to occur, the total initial kinetic energy in center-of-mass frame must exceed $\delta m_X$.  This
yields the constraint $(1/2) \mu_A v^2 > \delta m_X$, and as $m_A$ decreases, $v$ must increase in order for this
constraint to be satisfied.

We can instead parameterize the dark matter-nucleus differential scattering cross-section as
\bea
{d\sigma^{Z,A} \over dE_R} = {m_A I\over 32\pi v^2}
\left[Z + {f_n \over f_p}(A-Z)\right]^2
|F_A (E_R)|^2 ,
\label{InelasticScatteringEq}
\eea
where the squared dark matter-nucleus scattering matrix element can be written as
$m_X^2 m_A^2 [Z + (A-Z)(f_n/ f_p)]^2 |F_A (E_R)|^2 \times I$.  $I$ is a quantity which,
at leading order, is thus independent of the target nucleus mass and the relative velocity.

In figure~\ref{fig:Inelastic}~\cite{Kumar:2012uh} we plot the dark matter capture rate as a function of $m_X$ (in the low-mass regime)
for $I / 32\pi = 10^{-4}~\pb ~\gev^{-2}$, for each element in the sun.  As expected, we see that, as $\delta m_X$
increases, the capture rate from scattering off light elements decreases.  Interestingly,
we find that the capture is significant even for relatively large $\delta m_X$.  Indeed, dark matter
capture in the sun is possible even for values of $\delta m_X$ which would render inelastic scattering
on earth kinematically impossible.  The reason is that dark matter gains significant kinetic energy
as it falls towards the sun, allowing inelastic scattering even for large $\delta m_X$.  The
escape velocity at the surface of the sun is $\sim 600~\km / \s$; one would expect the range of $\delta m_X$
accessible to neutrino detectors via dark matter capture in the sun to be roughly a factor of 10 larger
than that accessible to earth-based direct detection experiments.

\begin{figure}[tb]
\includegraphics*[width=0.39\columnwidth]{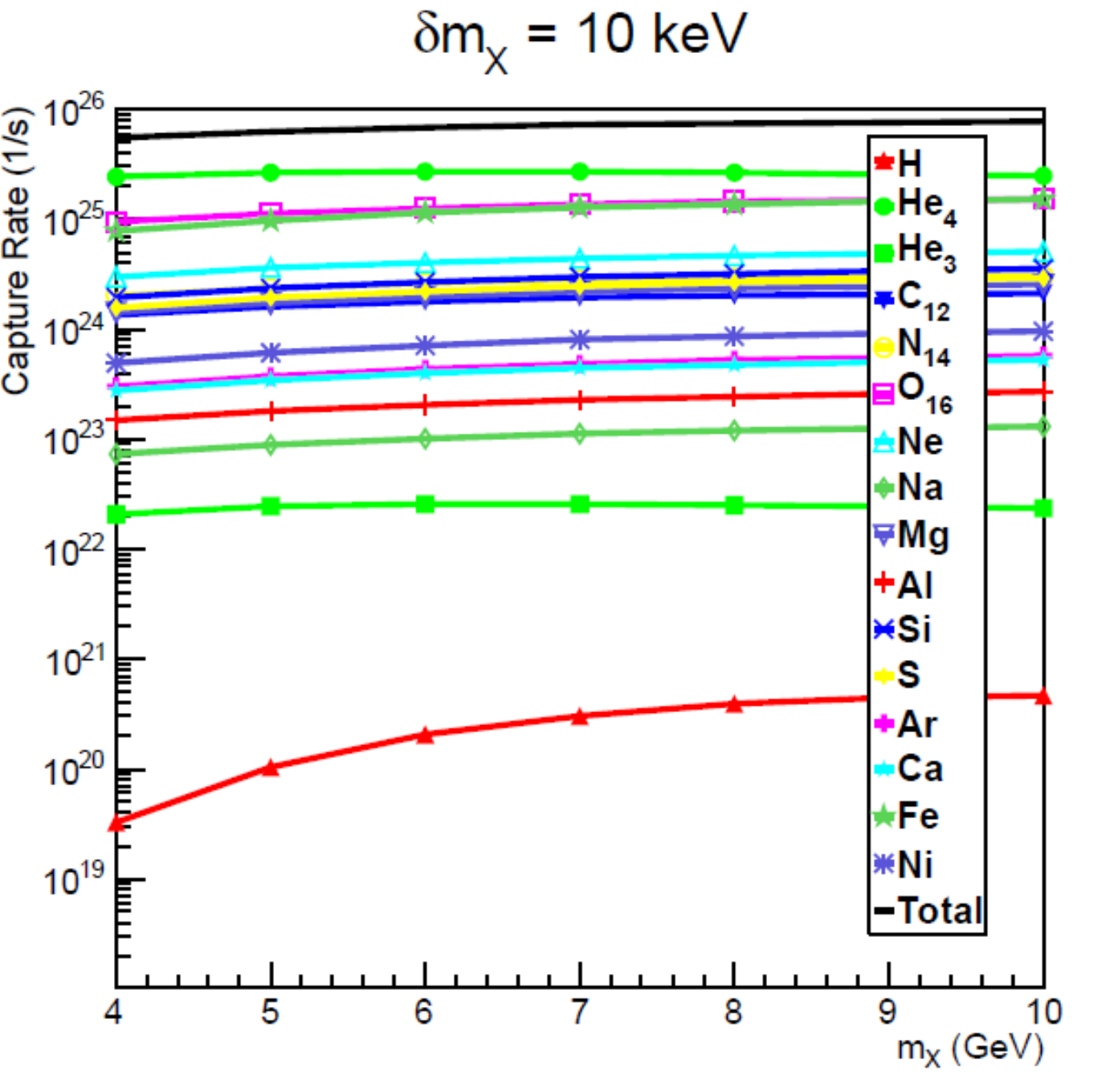}
\includegraphics*[width=0.4\columnwidth]{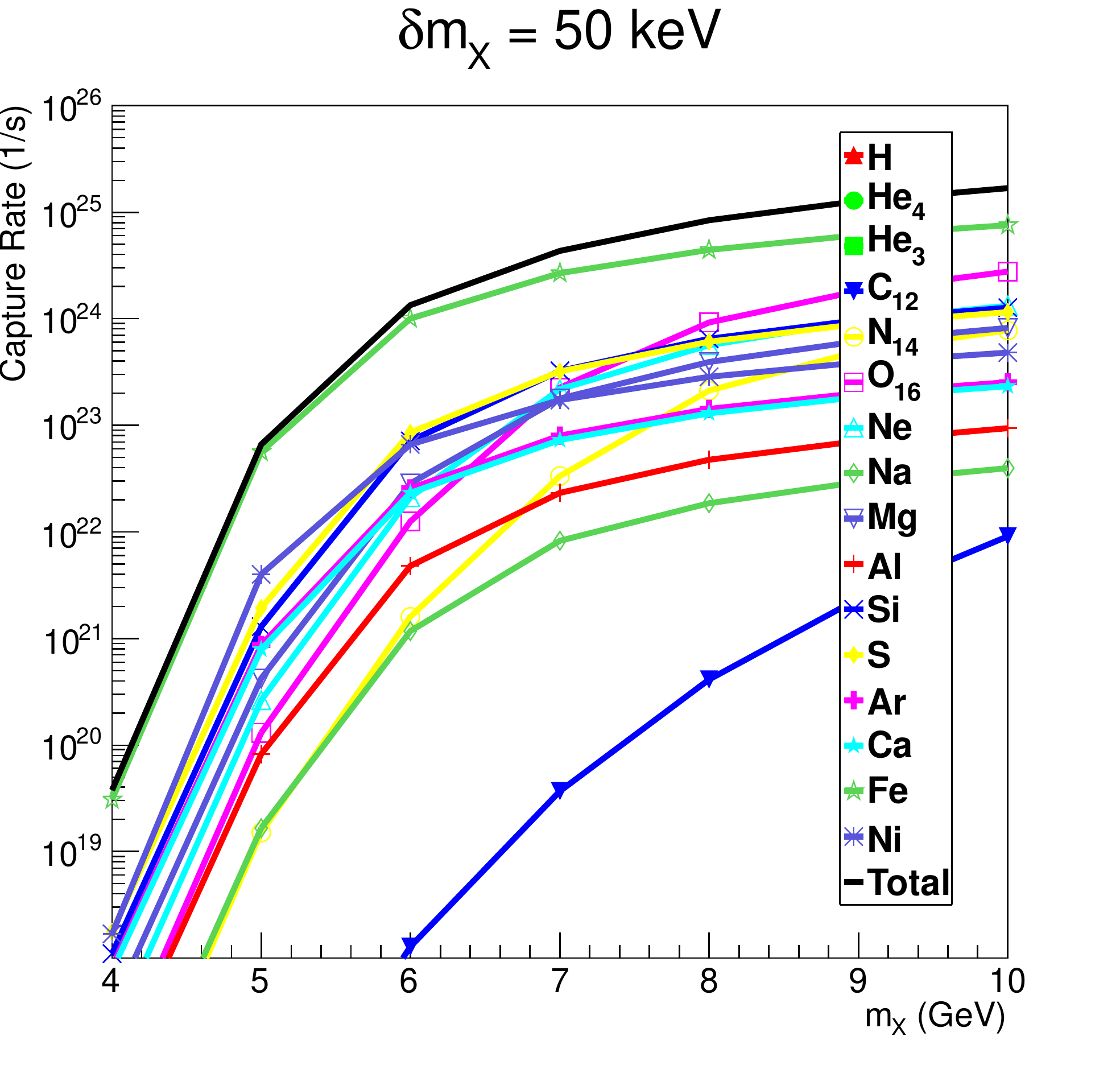}
\vspace*{-.1in}
\caption{\label{fig:Inelastic} Dark matter capture rates for various elements in the sun, assuming
isospin-invariant inelastic elastic contact interactions with $\delta m_X = 10~\kev$ (left panel) and
$50~\kev$ (right panel).  We set $I / 32\pi = 10^{-4}~\pb ~\gev^{-2}$.  }
\end{figure}


\section{Conclusion}

We have seen that, by relaxing some of the assumptions typically made regarding dark matter interactions,
one can dramatically change the interpretation of data from direct detection, indirect detection and
collider experiments.  This suggests that, to gain a definitive understanding of the properties of dark matter,
one may need to combine data from a variety of different detectors.  Moreover, one finds that some reasonable
dark matter models are best probed with novel detection strategies.  Neutrino detectors, gamma-ray searches
of dwarf spheroidal galaxies, and collider searches can all play an important complementary role.

All of these new possibilities for dark matter interactions have interesting
implications for upcoming detectors.  An illustrative example is the ${\rm D^3}$ experiment~\cite{Vahsen:2011qx},
which is under construction at the University of Hawaii.  This experiment is a low-threshold gas time projection
chamber which uses ionization to image the track made by a recoiling gas nucleus.  An initial plan
is for this detector to use a fluorocarbon target.  One can thus see that, for IVDM models with
destructive interference, detectors such as ${\rm D^3}$ can have enhanced sensitivity as compared to
other direct detection experiments.  For example, one can compare the normalized-to-nucleon cross-section
which would be seen by a fluorine-based detector to that of a germanium-based detector if
$f_n / f_p =-0.7$, in which case one finds $\sigma_N^{Z=F} = 4.2 \times \sigma_N^{Z=Ge}$.
Another interesting possibility is the use of a hydrocarbon target.  This could be especially
interesting in the case where dark matter interactions are long-ranged.  As we have seen, the
squared scattering matrix element scales as  $m_A^{-2}$ in the case of long-range interactions,
implying that a detector with a hydrogen target would receive an ${\cal O}(10^3)$ enhancement in
sensitivity relative to a
detector with a germanium target.

As more data
comes in from a variety of current and upcoming experiments, it will be important to keep in mind the effect
which relaxing typical theoretical assumptions can have on a coherent interpretation of the data.


\begin{theacknowledgments}
We are grateful to J.~L.~Feng, Y.~Gao, D.~Marfatia, K.~Richardson, M.~Sakai, D.~Sanford, S.~Smith and
L.~Strigari, for collaboration.  We are grateful to HOSC and 
we thank the Center for Theoretical Underground Physics and Related Areas (CETUP* 2012) in South Dakota 
for its hospitality and for partial support during the completion of this work.
This work is supported in
part by the Department of Energy under Grant~DE-FG02-04ER41291.
\end{theacknowledgments}



\bibliographystyle{aipproc}   

\bibliography{sample}

\begin{thebibliography}{9}

\bibitem{Goodman:1984dc}
  M.~W.~Goodman and E.~Witten,
  Phys.\ Rev.\ D {\bf 31}, 3059 (1985).

\bibitem{DDM}
  K.~R.~Dienes and B.~Thomas,
  Phys.\ Rev.\ D {\bf 85}, 083523 (2012)
  [arXiv:1106.4546 [hep-ph]];
  K.~R.~Dienes and B.~Thomas,
  Phys.\ Rev.\ D {\bf 85}, 083524 (2012)
  [arXiv:1107.0721 [hep-ph]];
  K.~R.~Dienes and B.~Thomas,
  Phys.\ Rev.\ D {\bf 86}, 055013 (2012)
  [arXiv:1203.1923 [hep-ph]];
  K.~R.~Dienes, S.~Su and B.~Thomas,
  Phys.\ Rev.\ D {\bf 86}, 054008 (2012)
  [arXiv:1204.4183 [hep-ph]];
  K.~R.~Dienes, J.~Kumar and B.~Thomas,
  Phys.\ Rev.\ D {\bf 86}, 055016 (2012)
  [arXiv:1208.0336 [hep-ph]].

\bibitem{Gould:1987ir}
  W.~H.~Press and D.~N.~Spergel,
  Astrophys.\ J.\  {\bf 296}, 679 (1985);
  A.~Gould,
  Astrophys.\ J.\  {\bf 321}, 571 (1987).


\bibitem{ScalarExamples}
  N.~G.~Deshpande and E.~Ma,
  Phys.\ Rev.\ D {\bf 18}, 2574 (1978);
  J.~L.~Feng, J.~Kumar and L.~E.~Strigari,
  Phys.\ Lett.\ B {\bf 670}, 37 (2008)
  [arXiv:0806.3746 [hep-ph]];
  S.~Andreas, T.~Hambye and M.~H.~G.~Tytgat,
  JCAP {\bf 0810}, 034 (2008)
  [arXiv:0808.0255 [hep-ph]].


\bibitem{olderIVDM}
  A.~Kurylov and M.~Kamionkowski,
  Phys.\ Rev.\  D {\bf 69}, 063503 (2004)
  [arXiv:hep-ph/0307185];
  F.~Giuliani,
  Phys.\ Rev.\ Lett.\  {\bf 95}, 101301 (2005)
  [arXiv:hep-ph/0504157];
  S.~Chang {\it et al.},
  JCAP {\bf 1008}, 018 (2010)
  [arXiv:1004.0697 [hep-ph]];
  Z.~Kang,
  {\it et al.},
  JCAP {\bf 1101}, 028 (2011)
  [arXiv:1008.5243 [hep-ph]].


\bibitem{Feng:2011vu}
  J.~L.~Feng,
  {\it et al.},
  Phys.\ Lett.\ B {\bf 703}, 124 (2011)
  [arXiv:1102.4331 [hep-ph]].

\bibitem{Bernabei:2010mq}
  R.~Bernabei,
  {\it et al.},
  Eur.\ Phys.\ J.\  {\bf C67}, 39-49 (2010).
  [arXiv:1002.1028 [astro-ph.GA]].

\bibitem{DAMAregion}
  C.~Savage {\it et al.}
  JCAP {\bf 0904}, 010 (2009)
  [arXiv:0808.3607 [astro-ph]];
  C.~Savage {\it et al.},
  Phys.\ Rev.\ D {\bf 83}, 055002 (2011)
  [arXiv:1006.0972 [astro-ph.CO]].


\bibitem{Aalseth:2010vx}
  C.~E.~Aalseth {\it et al.},
  Phys.\ Rev.\ Lett.\  {\bf 106}, 131301 (2011)
  [arXiv:1002.4703 [astro-ph.CO]].  

\bibitem{Aalseth:2011wp}
  C.~E.~Aalseth {\it et al.},
  Phys.\ Rev.\ Lett.\  {\bf 107}, 141301 (2011)
  [arXiv:1106.0650 [astro-ph.CO]].

\bibitem{Angloher:2011uu}
  G.~Angloher
  {\it et al.},
  [arXiv:1109.0702 [astro-ph.CO]].

\bibitem{Akerib:2010pv}
  D.~S.~Akerib {\it et al.},
  Phys.\ Rev.\ D {\bf 82}, 122004 (2010)
  [arXiv:1010.4290 [astro-ph.CO]].

\bibitem{Ahmed:2010wy}
  Z.~Ahmed {\it et al.},
  Phys.\ Rev.\ Lett.\  {\bf 106}, 131302 (2011)
  [arXiv:1011.2482 [astro-ph.CO]].

\bibitem{Angle:2011th}
  J.~Angle {\it et al.},
  Phys.\ Rev.\ Lett.\  {\bf 107}, 051301 (2011)
  [arXiv:1104.3088 [astro-ph.CO]].  

\bibitem{Aprile:2010um}
  E.~Aprile {\it et al.},
  Phys.\ Rev.\ Lett.\  {\bf 105}, 131302 (2010)
  [arXiv:1005.0380 [astro-ph.CO]].

\bibitem{Aprile:2011hi}
  E.~Aprile {\it et al.},
  Phys.\ Rev.\ Lett.\  {\bf 107}, 131302 (2011)
  [arXiv:1104.2549 [astro-ph.CO]].

\bibitem{Aprile:2012nq}
  E.~Aprile {\it et al.}  [XENON100 Collaboration],
  arXiv:1207.5988 [astro-ph.CO].


\bibitem{Felizardo:2011uw}
  M.~Felizardo,
  {\it et al.},
  [arXiv:1106.3014 [astro-ph.CO]].

\bibitem{Behnke:2012ys}
  E.~Behnke {\it et al.}  [COUPP Collaboration],
  Phys.\ Rev.\ D {\bf 86}, 052001 (2012)
  [arXiv:1204.3094 [astro-ph.CO]];

\bibitem{Aalseth:2012if}
  C.~E.~Aalseth {\it et al.}  [CoGeNT Collaboration],
  arXiv:1208.5737
  [astro-ph.CO].

\bibitem{Gao:2011bq}
  Y.~Gao, J.~Kumar and D.~Marfatia,
  Phys.\ Lett.\ B {\bf 704}, 534 (2011)
  [arXiv:1108.0518 [hep-ph]].

\bibitem{bib:xenon100}
See talk by D. Cline, {\tt http://public.lanl.gov/friedland/info11/info11talks/ClineDM-\\INFO11.pdf}



\bibitem{SuperCDMSlimit}
See talk by T.~Saab,
{\tt http://indico.in2p3.fr/contributionDisplay.py?sessionId=26\&\\contribId=58\&confId=1565}

\bibitem{DEAPCLEAN}
See talk by R.~Hennings-Yeomans,
{\tt http://deapclean.org/talks/PHENO2011\_Hennings.pdf}

\bibitem{bib:ic80dc}
  C.~d.~l.~Heros [for the IceCube Collaboration],
  arXiv:1012.0184 [astro-ph.HE].

\bibitem{LowMassNeutrinoDetector}
  D.~Hooper,
  {\it et al.},
  Phys.\ Rev.\ D {\bf 79}, 015010 (2009)
  [arXiv:0808.2464 [hep-ph]];
  J.~L.~Feng,
  {\it et al.},
  JCAP {\bf 0901}, 032 (2009)
  [arXiv:0808.4151 [hep-ph]];
  J.~Kumar, J.~G.~Learned and S.~Smith,
  Phys.\ Rev.\ D {\bf 80}, 113002 (2009)
  [arXiv:0908.1768 [hep-ph]];
  A.~L.~Fitzpatrick, D.~Hooper and K.~M.~Zurek,
  Phys.\ Rev.\ D {\bf 81}, 115005 (2010)
  [arXiv:1003.0014 [hep-ph]];
  S.~-L.~Chen and Y.~Zhang,
  Phys.\ Rev.\ D {\bf 84}, 031301 (2011)
  [arXiv:1106.4044 [hep-ph]].



\bibitem{Kumar:2011hi}
  J.~Kumar,
  {\it et al.},
  Phys.\ Rev.\ D {\bf 84}, 036007 (2011)
  [arXiv:1103.3270 [hep-ph]].  


\bibitem{Kumar:2012uh}
  J.~Kumar,
  {\it et al.},
  arXiv:1204.5120 [hep-ph].



\bibitem{LSDetector}
  J.~G.~Learned,
  arXiv:0902.4009 [hep-ex];
  J.~Peltoniemi,
  [arXiv:0909.4974 [physics.ins-det]].


\bibitem{Ackermann:2011wa}
  M.~Ackermann {\it et al.}  [Fermi-LAT Collaboration],
  Phys.\ Rev.\ Lett.\  {\bf 107}, 241302 (2011)
  [arXiv:1108.3546 [astro-ph.HE]].

\bibitem{GeringerSameth:2011iw}
  A.~Geringer-Sameth and S.~M.~Koushiappas,
  Phys.\ Rev.\ Lett.\  {\bf 107}, 241303 (2011)
  [arXiv:1108.2914 [astro-ph.CO]].


\bibitem{Kumar:2011dr}
  J.~Kumar, D.~Sanford and L.~E.~Strigari,
  Phys.\ Rev.\ D {\bf 85}, 081301 (2012)
  [arXiv:1112.4849 [astro-ph.CO]].

\bibitem{arXiv:1110.4376}
  K.~Abe
  {\it et al.},
  Phys.\ Rev.\ Lett.\  {\bf 108}, 051102 (2012)  [arXiv:1107.6000 [astro-ph.HE]];
  R.~Kappl and M.~W.~Winkler,
  arXiv:1110.4376 [hep-ph].

\bibitem{Evoli:2011id}
  C.~Evoli,
  {\it et al.},
  Phys.\ Rev.\ D {\bf 85}, 123511 (2012)
  [arXiv:1108.0664 [astro-ph.HE]].


\bibitem{Aad:2011xw}
  ATLAS Collaboration, ATLAS-CONF-2011-096.

\bibitem{MonojetCollider}
  S.~Chatrchyan {\it et al.}  [CMS Collaboration],
  Phys.\ Rev.\ Lett.\  {\bf 107}, 201804 (2011)
  [arXiv:1106.4775 [hep-ex]];
  T.~Aaltonen {\it et al.}  [CDF Collaboration],
  Phys.\ Rev.\ Lett.\  {\bf 108}, 211804 (2012)
  [arXiv:1203.0742 [hep-ex]].




\bibitem{monojet}
  J.~L.~Feng, S.~Su, F.~Takayama,
  Phys.\ Rev.\ Lett.\  {\bf 96}, 151802 (2006)
  [hep-ph/0503117];
  J.~Goodman, {\it et al.},
  Phys.\ Lett.\  {\bf B695}, 185-188 (2011)
  [arXiv:1005.1286 [hep-ph]];
  Y.~Bai, P.~J.~Fox and R.~Harnik,
  JHEP {\bf 1012}, 048 (2010)
  [arXiv:1005.3797 [hep-ph]];
  J.~Wang,
  {\it et al.},
  Phys.\ Rev.\ D {\bf 84}, 075011 (2011)
  [arXiv:1107.2048 [hep-ph]];  
  A.~Rajaraman, {\it et al.},
  [arXiv:1108.1196 [hep-ph]];
  J.~Goodman, {\it et al.},
  Phys.\ Rev.\  {\bf D82}, 116010 (2010)
  [arXiv:1008.1783 [hep-ph]];
  P.~J.~Fox,
  {\it et al.},
  Phys.\ Rev.\ D {\bf 85}, 056011 (2012)
  [arXiv:1109.4398 [hep-ph]];
  J.~Goodman, W.~Shepherd,
  [arXiv:1111.2359 [hep-ph]].

\bibitem{Akiri:2011dv}
  T.~Akiri {\it et al.}  [LBNE Collaboration],
  arXiv:1110.6249 [hep-ex].

\bibitem{iDM}
  D.~Tucker-Smith and N.~Weiner,
  Phys.\ Rev.\ D {\bf 64}, 043502 (2001)
  [hep-ph/0101138];
  D.~Tucker-Smith and N.~Weiner,
  Phys.\ Rev.\ D {\bf 72}, 063509 (2005)
  [hep-ph/0402065].



\bibitem{Vahsen:2011qx}
  S.~E.~Vahsen, 
  {\it et al.},
  arXiv:1110.3401
  [astro-ph.IM].



\end{thebibliography}

\IfFileExists{\jobname.bbl}{}
 {\typeout{}
  \typeout{******************************************}
  \typeout{** Please run "bibtex \jobname" to optain}
  \typeout{** the bibliography and then re-run LaTeX}
  \typeout{** twice to fix the references!}
  \typeout{******************************************}
  \typeout{}
 }

\end{document}


\endinput